\documentclass[showpacs,prl,preprintnumbers,amsmath,amssymb]{revtex4}

\usepackage{graphicx}
\usepackage{dcolumn}
\usepackage{bm}

\begin{document}

\title{Nonequilibrium self-gravitational system}

\author{B. I. Lev}

\affiliation{Bogolyubov Institute for Theoretical Physics, NAS of
Ukraine, Metrolohichna 14-b, Kyiv 03680, Ukraine}

\date{\today}

\begin{abstract}

The density of states of self-gravitational system diverges when
the particles are spread to infinity. Other problem based an
inhomogeneous distribution of particles,which motivate the
gravitational interaction. In this sense the statistical
mechanics of self-gravitational system is essentially an
non-equilibrium problem. A new possible approach to statistical
description of self-gravitational system has been proposed. The
approach based on non-equilibrium statistical operator, which
allow take into account inhomogeneous distribution particle and
temperature in self gravitational system. The saddle point
procedure, which used in the given method describes the spatially
inhomogeneous distribution in self gravitational system
accompanied by temperature changing.
\end{abstract}

\maketitle

The statistical description of a system of interacting particles
is a difficult problem but has the permanent attention for the
problems of astrophysics \cite{Sas}.The self-gravitational system
are interesting for testing ideas about the statistical mechanic
description of systems governed by long range interaction. The
statistical mechanics of self gravitational system turns out to
be very different from that of other, more familiar, many-body
systems. A self-gravitational system has also more general
problem, studied for a long time \cite{Pad}. The standard methods
of statistical mechanic cannot be carrier to study gravitational
system. Due to this fundamental difference, the notion of
equilibrium is not always well defined and those system exhibit a
nontrivial behaviour with gravitational collapse. For system with
long-range interaction the thermodynamically ensemble are
inequivalent, negative specific heat \cite{Tir} in microcanonical
ensemble which not exist in canonical description \cite{Cha}.In
self-gravitational system can increase entropy and equilibrium
states are only local entropy maximum. However, if introduce a
repulsive potential at short distance, complete core collapse is
prevented and can to proved that a global entropy maximum now
exist for all accessible values of energy \cite{Aro}. For the gas
with pure gravitational interaction between its particles the
partition function diverges, the energy of self-gravitational
system are not as extensive parameter. In it was shown that a
self-gravitating gas collapses. A nature of the collapse and its
conditions are explained by using simple and clear consideration
\cite{Lyn}. The phase transition in such systems creates the
problem of description in the mean-field thermodynamics approach
\cite{Cha}. Two type approaches (statistical and thermodynamic)
have been develop to determination the equilibrium states of
self-gravitational system \cite{Cha}, \cite{Cav}. About all this
problem and possible solution very good described in review
\cite{Chav}. The collapse in such system are begin as spatially
homogeneous distribution of particles in the all system at once.
Formation of the spatially inhomogeneous distribution of
interaction particles is a typical problem in condensed matter
physics and requires non-conventional methods of statistical
description of the system was tailored to gravitational
interacting particles with regard for an arbitrary spatially
inhomogeneous particle distribution. Spatially inhomogeneous
particle distribution can not describe in standard approach
equilibrium statistical physics. The microcanonical ensemble is
the correct description of an isolated Hamiltonian self
gravitational system at fixed energy, the canonical ensemble is
the correct description of a dissipative Brownian system at fixed
temperature \cite{Chav}. For inhomogeneous distribution of
particle can not fixed any energy or temperature. Temperature can
change together with distribution of particle in particular if
fixed energy of system. Very important question are in point can
be considered inhomogeneous distribution of particle as
equilibrium state. This state is not stationary and derivation of
density produce flux of particle though changing of gravitational
force which act on every particle. The statistical description of
self-gravitational system must take into account the possible
inhomogeneous distribution the temperature together with
distribution of particles. A few model systems with interaction
are known for which the partition function can be exactly
evaluated, at least within thermodynamic limits \cite{Bax} but
not for inhomogeneous distribution of particle \cite{Aly}.
Inhomogeneous distribution of particle and temperature need to
use nonequilibrium approach to description of behaviour self
gravitational system. In this article present the possible
approach local equilibrium statistical operator \cite{Zub} which
leads to take into account the inhomogeneous distribution
particle and temperature. Formation of the spatially inhomogeneous
distribution of interaction particles requires a nonconventional
method, such as use in \cite{Bel}, \cite{Lev}, \cite{Kle}, which
are based an additional field representation of statistical sum
\cite{Str}. For progress the goal can use saddle point
approximation which provided to nonlinear equation. Solution of
this equation present real distribution particle and temperature
under certain conditions. In this approach do not use the
polytrophic dependence pressure from density. This dependence are
result as consequence of the thermodynamic relation. Presented
condition give the possibility determine statistical operator for
self gravitational system and fully describe the thermodynamic
behaviour of self-gravitational system.

Phenomenological thermodynamic based on the conservation lows for
average value of physical parameter as number of particles,
energy and impulse. Statistical thermodynamic nonequilibrium
system based too on conservation lows not the average value
dynamic variables but in particular for this dynamic variables.
For determination thermodynamic function of nonequilibrium system
are need use the presentation of corresponding statistical
ensembles which take into account the nonequilibrium states of
this systems. The conception of Gibbs ensembles can brings to
description nonequilibrium stationary states of system. In this
case can determine nonequilibrium ensemble as totality of system
which be contained in same stationary external action. This
system have same character of contact with thermostat and possess
all possible value macroscopical parameters which compatibility
present conditions. In system, which are in same stationary
external condition will be formed local equilibrium stationary
distribution. For exactly determination local equilibrium
ensemble must accordingly determine the distribution function or
statistical operator of system \cite{Zub}.

If assume that nonequilibrium states of system can determine
through inhomogeneous distribution energy $H(\mathbf{r})$ and
number of particles (density
$n(\mathbf{r})=\sum_{i}\delta(\mathbf{r}-\mathbf{r_{i}})$ ) the
local equilibrium distribution function for classical system can
write in the form \cite{Zub}:$f_{l}=Q^{-1}_{l}\exp\left\{-\int
\left(\beta(\mathbf{r})H(\mathbf{r})-\eta(\mathbf{r})n(\mathbf{r})\right)d\mathbf{r}\right\}$
where statistical operator local equilibrium distribution can
determine as:
\begin{equation}
Q_{l}=\int D\Gamma \exp\left\{-\int
\left(\beta(\mathbf{r})H(\mathbf{r})-\eta(\mathbf{r})n(\mathbf{r})\right)d\mathbf{r}\right\}
\end{equation}
The integration in present formula must take over all phase space
of system. Must note, that in the case local equilibrium
distribution Lagrange multipliers $\beta(\mathbf{r})$ and
$\eta(\mathbf{r})$ are function of spatial point. Can introduce
the entropy local equilibrium distribution by ratio
\begin{equation}
S=-Sp\left(f_{l}\ln f_{l}\right)=\ln Q_{l}+ \int
\left(\beta(\mathbf{r})H(\mathbf{r})-\eta(\mathbf{r})n(\mathbf{r})\right)d\mathbf{r}
\end{equation}
After determination of statistical operator can obtain all
thermodynamic parameter nonequilibrium system. For this, can
determine thermodynamic relation for the systems. The variation
of statistical operator by Lagrange multipliers can write
necessary thermodynamic relation in the form:
\begin{equation}
-\frac{\delta \ln Q_{l}}{\delta \beta(\mathbf{r})}=\langle
H(\mathbf{r})\rangle_{l}-\eta(\mathbf{r})\langle
n(\mathbf{r})\rangle_{l}
\end{equation}
and
\begin{equation}
\frac{\delta \ln Q_{l}}{\delta \eta(\mathbf{r})}=\langle
n(\mathbf{r})\rangle_{l}
\end{equation}
This relation is natural general prolongation, on the case
inhomogeneous system, well-known relation which take place in the
case equilibrium systems. The conservation number of particles
and energy in system can present in form natural relations $\int
n(\mathbf{r})d\mathbf{r}=N$ and $\int H(\mathbf{r})d\mathbf{r}=E$.

For further statistical description of nonequilibrium system is
necessary determine Hamiltonian of system. In the case of self
gravitational system Hamiltonian can present in the form:
\begin{equation}
H(\mathbf{r})=\frac{p^{2}(\mathbf{r})}{2m}n(\mathbf{r})+\frac{1}{2}\int
W(\mathbf{r},\mathbf{r'})n(\mathbf{r})n(\mathbf{r'})d\mathbf{r'}
\end{equation}
where impulse density
$p(\mathbf{r})=\sum_{i}p_{i}\delta(\mathbf{r}-\mathbf{r_{i}})$
and gravitation interaction energy can present in well-known form
\begin{equation}
W(\mathbf{r},\mathbf{r'})=\frac{G m^{2}}{|\mathbf{r}-\mathbf{r'}|}
\end{equation}
$G$ is gravitational constant and $m$ is mass of particle. This
Hamiltonian of system is possible to use if take into account
consider moving in phase space not compressed gravitational
fluid. It is valid for not collision systems and self
gravitational system  present obvious example of such system.

In the case self gravitational system the nonequilibrium
statistical operator can write in the form:
\begin{equation}
Q_{l}=\int D\Gamma \exp\left\{-\int
\left(\beta(\mathbf{r})\frac{p^{2}(\mathbf{r})}{2m}-\eta(\mathbf{r})\right)n(\mathbf{r})d\mathbf{r}-\frac{1}{2}\int
W(\mathbf{r},\mathbf{r'})n(\mathbf{r})n(\mathbf{r'})d\mathbf{r}d\mathbf{r'}\right\}
\end{equation}
In order to perform a formal integration in second part of this
presentation, additional field variables can be introduced making
use of the theory of Gaussian integrals \cite{Str}, \cite{Kle}:
\begin{equation}
\exp \left\{ -\frac{1}{2}\int
\beta(\mathbf{r})W(\mathbf{r},\mathbf{r'})n(\mathbf{r})n(\mathbf{r'})d\mathbf{r}d\mathbf{r'}\right\}
= \int D\varphi \exp \left\{ -\frac{1 }{2}\int
\beta(\mathbf{r})W^{-1}(\mathbf{r},\mathbf{r'})\varphi(\mathbf{r})\varphi(\mathbf{r'})d\mathbf{r}d\mathbf{r'}-
\int\sqrt{\beta(\mathbf{r})}\varphi(\mathbf{r})n(\mathbf{r})\right\}
\end{equation}
where $D\varphi=\frac{\prod\limits_{s}d\varphi _{s} }{\sqrt{\det
2\pi \beta W(\mathbf{r},\mathbf{r'})}}$ and
$W^{-1}(\mathbf{r},\mathbf{r'})$ is the inverse operator which
satisfies the condition $\int d\mathbf{r'}
W^{-1}(\mathbf{r},\mathbf{r'})W(\mathbf{r'},\mathbf{r''})=\delta
(\mathbf{r}-\mathbf{r''})$. The inverse operator
$W^{-1}(\mathbf{r},\mathbf{r'})$ of the gravitational interaction,
in continuum limit should be treated in the operator sense
\cite{Mag}, i.e.
\begin{equation}
W^{-1}(\mathbf{r},\mathbf{r'})=-\frac{1}{4\pi G
m^{2}}\triangle_{\mathbf{r}}\delta(\mathbf{r}-\mathbf{r'})
\end{equation}
where $\triangle_{\mathbf{r}}$- is Laplace operator in real space.
After this manipulation the statistical operator can rewrite in
the form:
\begin{equation}
Q_{l}=\int D\Gamma \int D\varphi \exp\left\{-\int
\left(\beta(\mathbf{r})\frac{p^{2}(\mathbf{r})}{2m}-\eta(\mathbf{r})-\sqrt{\beta(\mathbf{r})}\right)n(\mathbf{r})d\mathbf{r}-
\frac{1}{8\pi m^{2}G}\int \left(\nabla \varphi(\mathbf{r})
\right)^{2}d\mathbf{r}\right\}
\end{equation}
In this functional integral now can be provide the integration on
phase space. The integration over phase space can present as
functional integration by
$D\Gamma=\frac{Dn(\mathbf{r})Dp(\mathbf{r})}{\left( 2\pi
\hbar\right) ^{3}}$ with regard for the cell volume $\left( 2\pi
\hbar\right) ^{3}$ in the phase space of individual states
\cite{Lev}, and take into account the "Pauli principle" for
classical system, that own spatial point can not occupier two
classical particle. As result can use the relation $\int
\exp(-A(\mathbf{r})n(\mathbf{r}))Dn(\mathbf{r})=\frac{1}{\det
A(\mathbf{r})}\left\{1-\exp\left(-\int
A(\mathbf{r})d\mathbf{r}\right)\right\}$. Now can make functional
integration over impulse. If introduce new variable absolute
chemical activity $\xi(\mathbf{r})\equiv \exp \eta(\mathbf{r})$
after integration over impulse can write statistical operator in
the form:
\begin{equation}
Q_{l}=\int D\varphi \exp\left\{-\int \left[\frac{1}{8\pi m^{2}G}
\left(\nabla \varphi(\mathbf{r})\right)^{2}- \ln\left(1-
\xi(\mathbf{r})(\frac{2\pi
m}{\hbar^{2}\beta(\mathbf{r})})^{\frac{3}{2}}\exp\sqrt{\beta(r)}\varphi(\mathbf{r})\right)\right]d\mathbf{r}\right\}
\end{equation}
As shown before \cite{Bel},\cite{Lev} in all cases classical
statistic $\xi\leq 1$ and can use expansion $\ln \left( 1-\xi A
\right)\approx -\xi A+...$ and rewrite the nonequilibrium
statistical operator in more simple form:
\begin{equation}
Q_{l}=\int D\varphi \exp\left\{\int \left[-\frac{1}{8\pi m^{2}G}
\left(\nabla \varphi(\mathbf{r})\right)^{2}+
\xi(\mathbf{r})\frac{2\pi
m}{\hbar^{2}\beta(\mathbf{r})})^{\frac{3}{2}}\exp\sqrt{\beta(r)}\varphi(\mathbf{r})\right]d\mathbf{r}\right\}
\end{equation}
In the case constant temperature $\beta$ and absolute chemical
activity $\xi$ the statistical operator fully reconstruct the
equilibrium grand canonical partition function \cite{Lev},
\cite{Veg}.

The statistical operator allows obtain use the of efficient
methods developed in the quantum field theory without imposing
additional restrictions of integration over field variables or
the perturbation theory. In our case the nonequilibrium
statistical operator can rewrite in the form
\begin{equation}
Q_{l}=\int D\varphi
\exp\left\{-F(\varphi(\mathbf{r}),\xi(\mathbf{r}),\beta(r))\right\}
\end{equation}
where effective nonequilibrium "free energy" can present as:
\begin{equation}
F(\varphi(\mathbf{r}),\xi(\mathbf{r}),\beta(r))=\int
\left[-\frac{1}{8\pi m^{2}G} \left(\nabla
\varphi(\mathbf{r})\right)^{2}+ \xi(\mathbf{r})\frac{2\pi
m}{\hbar^{2}\beta(\mathbf{r})})^{\frac{3}{2}}\exp\sqrt{\beta(r)}\varphi(\mathbf{r})\right]d\mathbf{r}
\end{equation}
The functional $F(\varphi(\mathbf{r}),\xi(\mathbf{r}),\beta(r))$
depends on distribution of the field variables
$\varphi(\mathbf{r} $, chemical activity $\xi(\mathbf{r}$ and
temperature $\beta(\mathbf{r})$. The field variable contains the
same information as original distribution, information about
possible states of the systems. The saddle point method can now
be further employed  can find the asymptotic value of the
statistical operator $Q_{l}$; the dominant contribution is given
by the states which satisfy the extreme condition for the
functional. It's easy to see that saddle point equation present
thermodynamic relation and it can write in the other form: as
equation for field variable $\frac{\delta F}{\delta
\varphi(\mathbf{r})}=0$, the normalization condition $ \int
d\mathbf{r}\frac{\delta F}{\delta (\eta(\mathbf{r}))}=-\int
\frac{\delta F}{\delta
(\xi(\mathbf{r}))}\xi(\mathbf{r}))d\mathbf{r}=N$ and low of
conservation the energy of system $ -\int d\mathbf{r}\frac{\delta
F}{\delta (\beta(\mathbf{r}))}\xi(\mathbf{r})=E$. Solution of
this equation fully determine all thermodynamic parameter and
present general solution for the behaviour of self gravitational
system, whether this distribution of particles is spatially
inhomogeneous or not. The spatially inhomogeneous solution of
this equations corespondent the distribution of interacting
particles. Very important note, that only in this approach can
take into account the inhomogeneous distribution of temperature,
which can depend from spatial distribution of particle in system.
In other approaches the dependence of temperature from spatial
point was introduce trough polytrophic dependence temperature
from density of particle in equation of state. In present
approach this dependence leads from necessary thermodynamic
condition and can be determine for different distribution of
particles. From normalization condition $\int
\rho(\mathbf{r})d\mathbf{r}=N$ can introduce the density function
by definition in the form $\rho(\mathbf{r})\equiv \xi \frac{2\pi
m}{\hbar^{2}\beta(\mathbf{r})})^{\frac{3}{2}}\exp(\sqrt{\beta(\mathbf{r})}\varphi(\mathbf{r}))$
and rewrite the necessary equation in more simple presentation.
Equation for field variable can write in the form
\begin{equation}
\triangle
\varphi(\mathbf{r})+r_{m}\sqrt{\beta(\mathbf{r})}\rho(\mathbf{r})=0
\end{equation}
and equation of conservation energy take the form:
\begin{equation}
\int \frac{\rho(\mathbf{r})}{2\beta(\mathbf{r})}(3-
\sqrt{\beta(r)}\varphi(\mathbf{r}))d\mathbf{r}=E
\end{equation}
where $r_{m}\equiv 4\pi G m^{2}$. In the case absent any
interaction, from normalization condition $\int \xi \frac{2\pi
m}{\hbar^{2}\beta})^{\frac{3}{2}}dV=N$ at constant temperature
can obtain the the chemical activity $\xi=\frac{N}{V}\frac{2\pi
m}{\hbar^{2}\beta})^{-\frac{3}{2}}$ and from equation
conservation energy relation between energy and temperature
$\frac{3}{2}NT=E$. If use obtained relation can determine usual
entropy in the form $S=\ln \frac{N!}{V}\left(\frac{4\pi mE}{3 N
\hbar^{2}}\right)^{\frac{3}{2}}+\frac{3N}{2}$ that fully
reproduce entropy microcanonical distribution gaseous phase for
fixed number of particles and energy in system. In the case
homogeneous distribution of particle of self gravitational system
can take $\rho\equiv\frac{N}{V}=const$. If temperature $T$ is
constant, the equation for field variable $\varphi$ leads to
$\triangle \varphi+\frac{4\pi G m^{2}}{T}\rho=0$ and have
solution $\varphi = \frac{G m^{2}}{T}\frac{1}{r}$. Substitution
this solution in the equation for conservation energy $\frac{N}{2
V}\int^{R}_{0} (3-\frac{G m^{2}}{T}\frac{1}{r})r^{2}dr=E$ can
obtain for the gravitational gaseous phase in limiting volume of
size $R$can obtain that $\frac{3}{2}NT=E+\frac{3G m^{2}}{8\pi
R}$. This relation completely reproduce well-known relation
\cite{Chav} which was obtained by standard approach.

In general case the distribution of particle in self-
gravitational system are inhomogeneous. Inhomogeneous
distribution of particle motivate the long-range gravitational
interaction. Can consider other simple case. If suppose that
exist polynomial relation between density and temperature as
$\sqrt{\beta(\mathbf{r})}\rho(\mathbf{r})=A=const$ and density
change in accordance with not divergence of number of particle,
that possible if density descend as fourth time of distance from
center $\rho(r)=Cr^{-4}$ from normalized condition can obtain,
that $C=NR_{0}$ where $R_{0}$ is smallest distance to center where
classical particle create close packing structure. From equation
by field variable $\varphi(\mathbf{r})=\frac{r_{m}A}{r}$ that
directly reconstruct behavior of gravitational field. From this
condition temperature descend as fourth time of distance and can
write that $\sqrt{\beta(\mathbf{r})}=\frac{Ar^{4}}{C}$ or
$T=\frac{C}{A}r^{-8}$. From equation for energy can obtain
relation between introducing constants in the form
$\frac{C^{2}}{9AR^{9}_{0}}-\frac{A C r_{m}}{6 R^{6}_{0}}=E$. In
this simple case can obtain all necessary coefficient and spatial
dependence density and temperature for inhomogeneous distribution
particles of self gravitational system.

Indeed,present nonequilibrium statistical description tell only
possible dilute structure in self-gravitational system but not
describe meta stable states and tell nothing about time scales a
kinetic theory. The statistical operator have not any peculiarity
for different value of gravitational field. The problem of
description of the self-gravitational system of particles could
be solved with current approach which take into account the
inhomogeneous distribution of particles and temperature. For the
first time turn out well described the formation spatial
inhomogeneous distribution particle with accompanied by changing
temperature such distribution of interacting particles. More
over, the method used can be also applied for further development
of physics of self- gravitational and similar systems.

\end{document}